\documentclass{article}
\usepackage{spconf,amsmath,graphicx}
\usepackage{mathtools, amsfonts, url, cite, hyperref, subcaption, booktabs, tabularx}
\newcolumntype{Y}{>{\centering\arraybackslash}X}

\title{Stylebook: Content-dependent speaking style modeling \\ for any-to-any voice conversion using only speech data}
\name{Hyungseob Lim$^1$\thanks{* Work done during the internship at Qualcomm Technologies, Inc.}$^*$ \qquad Kyungguen Byun$^2$ \qquad Sunkuk Moon$^2$ \qquad Erik Visser$^2$}
\address{$^1$ Department of Electrical and Electronic Engineering, Yonsei University, Seoul, South Korea
       \\$^2$ Qualcomm Technologies, Inc., San Diego, California, USA}
\begin{document}
%
\maketitle
\begin{abstract}
\vspace{-5pt}
While many recent any-to-any voice conversion models succeed in transferring some target speech's style information to the converted speech, they still lack the ability to faithfully reproduce the speaking style of the target speaker.
In this work, we propose a novel method to extract rich style information from target utterances and to efficiently transfer it to source speech content without requiring text transcriptions or speaker labeling.
Our proposed approach introduces an attention mechanism utilizing a self-supervised learning (SSL) model to collect the speaking styles of a target speaker each corresponding to the different phonetic content.
The styles are represented with a set of embeddings called stylebook.
In the next step, the stylebook is attended with the source speech's phonetic content to determine the final target style for each source content.
Finally, content information extracted from the source speech and content-dependent target style embeddings are fed into a diffusion-based decoder to generate the converted speech mel-spectrogram.
Experiment results show that our proposed method combined with a diffusion-based generative model can achieve better speaker similarity in any-to-any voice conversion tasks when compared to baseline models, while the increase in computational complexity with longer utterances is suppressed.
\vspace{-3pt}
\end{abstract}

\begin{figure*}
    \centering
    \includegraphics[width=0.85\textwidth]{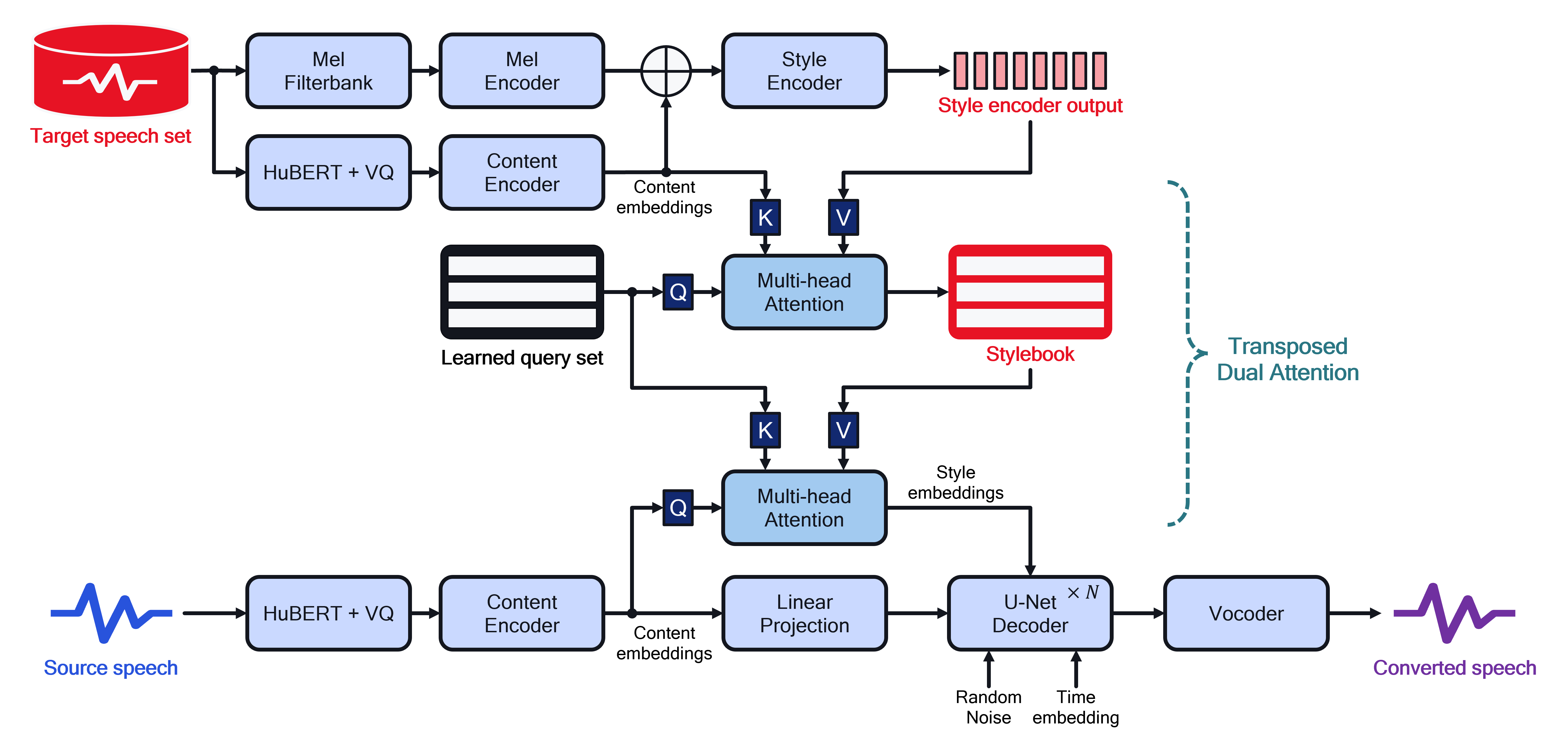}
    \vspace{-10pt}
    \caption{Overall structure of the proposed model for any-to-any voice conversion. Q, K, and V in the diagram refer to the attention heads for the queries, keys, and values, respectively. Once the stylebook (a set of style embeddings) is obtained from the target speaker and stored, the same computation does not have to be repeated for voice conversion.}
    \label{fig:model}
\vspace{-18pt}
\end{figure*}

\vspace{-8pt}
\section{Introduction}
\label{sec:intro}
\vspace{-8pt}
Voice conversion refers to the task of converting one person's voice (source) into another person's voice (target) while preserving linguistic content (e.g., phoneme, word) uttered by the source speaker.
Specifically, it is desired that the content of the converted speech is given by the content of source speech, while the speaking style (e.g., speaker identity, accent) resembles that of the target speaker.
Any-to-any voice conversion aims at coping with any unseen speaker's speech without any prior knowledge of the speakers.

A general approach for any-to-any voice conversion is to establish an autoencoder structure, which disentangles content-related information and style-related information at the encoder side and reconstructs the original signal at the decoder side using a deep generative network\cite{autovc, vqmivc, softvc, freevc, yourtts, diffvc}.
In such a framework, voice conversion is performed by first extracting content and style information from both the source speech and the target speech.
Then, the decoder takes the content information from the source and the style information from the target to generate the converted speech.
With regard to content information extraction, many recent works~\cite{vqmivc, softvc, freevc} utilize self-supervised learning (SSL) models \cite{wavlm, hubert} which do not require text transcription for training the content extractor.
However, most of the works still rely on a single, global vector (usually a speaker embedding extracted from a pre-trained speaker verification model) for conditioning all the frames, therefore failing to faithfully transfer the style of the target speaker to the source speech content.

On the other hand, kNN-VC \cite{knn-vc} (k-nearest neighbors voice conversion) recently proposed a simple but effective method to accomplish the content-preserving style transfer only with a pre-trained SSL model and a vocoder.
Motivated by the previous finding that SSL features extracted from a similar phonetic content lie in proximity in the embedding space regardless of the speaker identity \cite{vqmivc, softvc, wavlm}, kNN-VC finds kNNs for each source speech's WavLM \cite{wavlm} feature among the SSL features from the target speaker.
Then, a vocoder trained for synthesizing a waveform from the WavLM features is used to generate the converted speech from the averaged target WavLM features.
By doing so, the model achieved state-of-the-art performance when a long target speech (about 5 minutes) was given. However, the model suffers from increased computation time and memory usage for the long target speech due to the nature of the kNN search.

In this work, we propose a novel method to extract multiple style representations in a content-dependent manner without text transcription or speaker labeling and apply it to any-to-any voice conversion.
Using discretized SSL feature as disentangled phonetic content information as in \cite{vqmivc, softvc} along with the attention mechanism\cite{transformer}, our proposed model generates multiple style embeddings each responsible for the pronunciation of different phonetic content (hereafter called content-dependent style embeddings).
For the voice conversion, different style embeddings are extracted based on the source speech's content embeddings using another attention mechanism, and both embeddings are combined in a diffusion model \cite{score_diffusion} to generate converted speech.
To the best of our knowledge, it is the first approach that utilizes content-dependent style embeddings for conditioning a voice conversion system with no auxiliary losses.
Unlike the kNN-VC approach, our proposed model can efficiently capture the speaking style of a target speaker with a fixed-size array to avoid exhaustive computational processing in case of long target speech.
From experiments on any-to-any voice conversion with various length target speech, we show that our proposed model can achieve better speaker similarity than models utilizing single speaker embedding and that performance is improved with more target data without increasing memory usage during inference.

\vspace{-22pt}
\section{PROPOSED MODEL}
\label{sec:proposed}
\vspace{-8pt}
Fig. \ref{fig:model} shows the overall structure of the proposed model for any-to-any voice conversion.
First, a stylebook (a set of style embeddings) is generated from a target speech set (utterances of a target speaker).
Then, style embeddings that correspond to the content of a source speech are selected from the stylebook via the attention mechanism.
The combination of the content from the source and the style from the target is passed into a diffusion-based \cite{score_diffusion} U-Net\cite{u-net} decoder and a neural vocoder to yield converted speech.
The following sections explain how each module works in detail.

\begin{figure*}
    \centering
    \begin{subfigure}{0.50\textwidth}
        \centering
        \includegraphics[width=\textwidth]{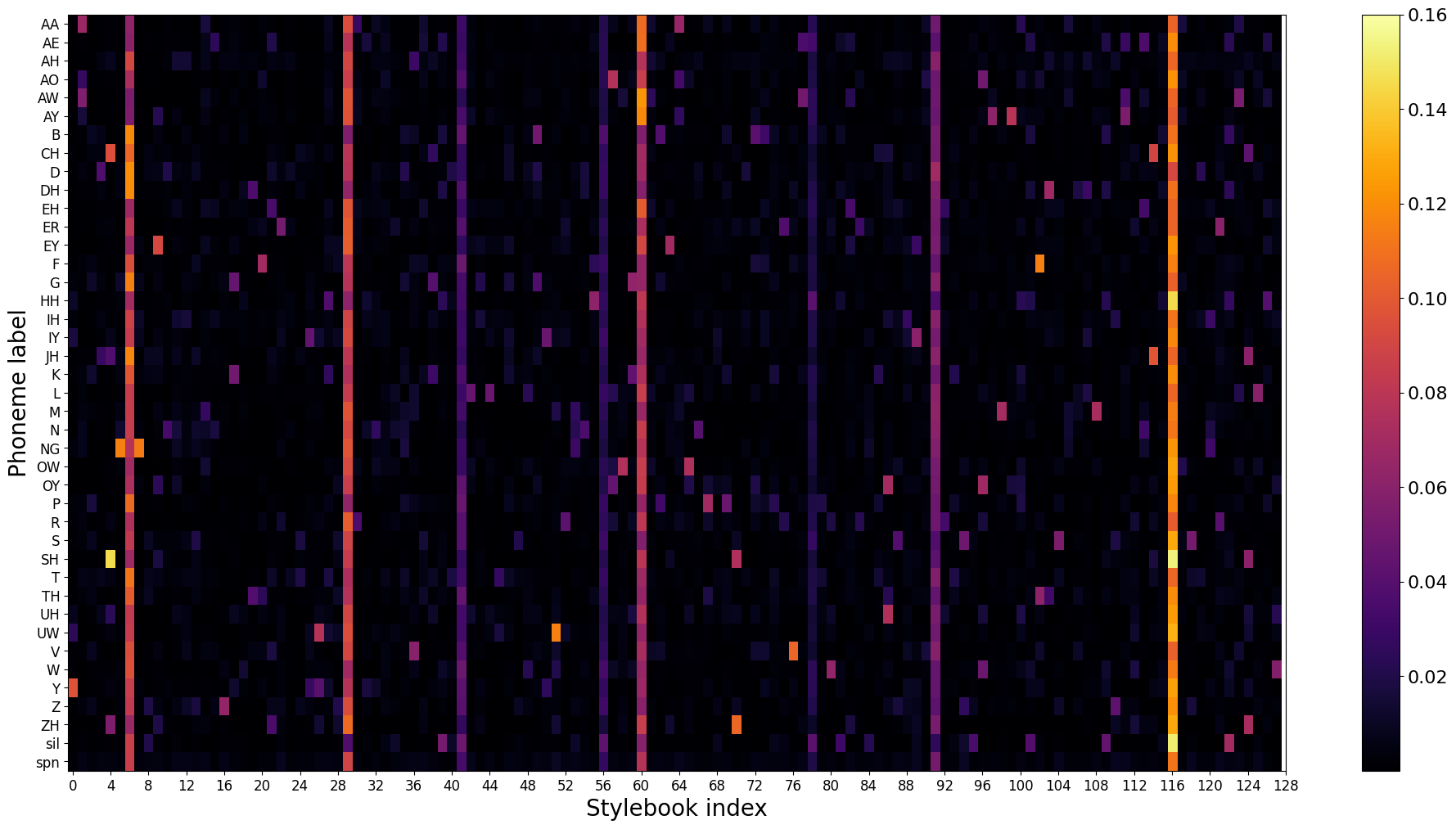}
        \caption{}
    \end{subfigure}
    \hspace{0.03\textwidth}
    \begin{subfigure}{0.355\textwidth}
        \centering
        \includegraphics[width=\textwidth]{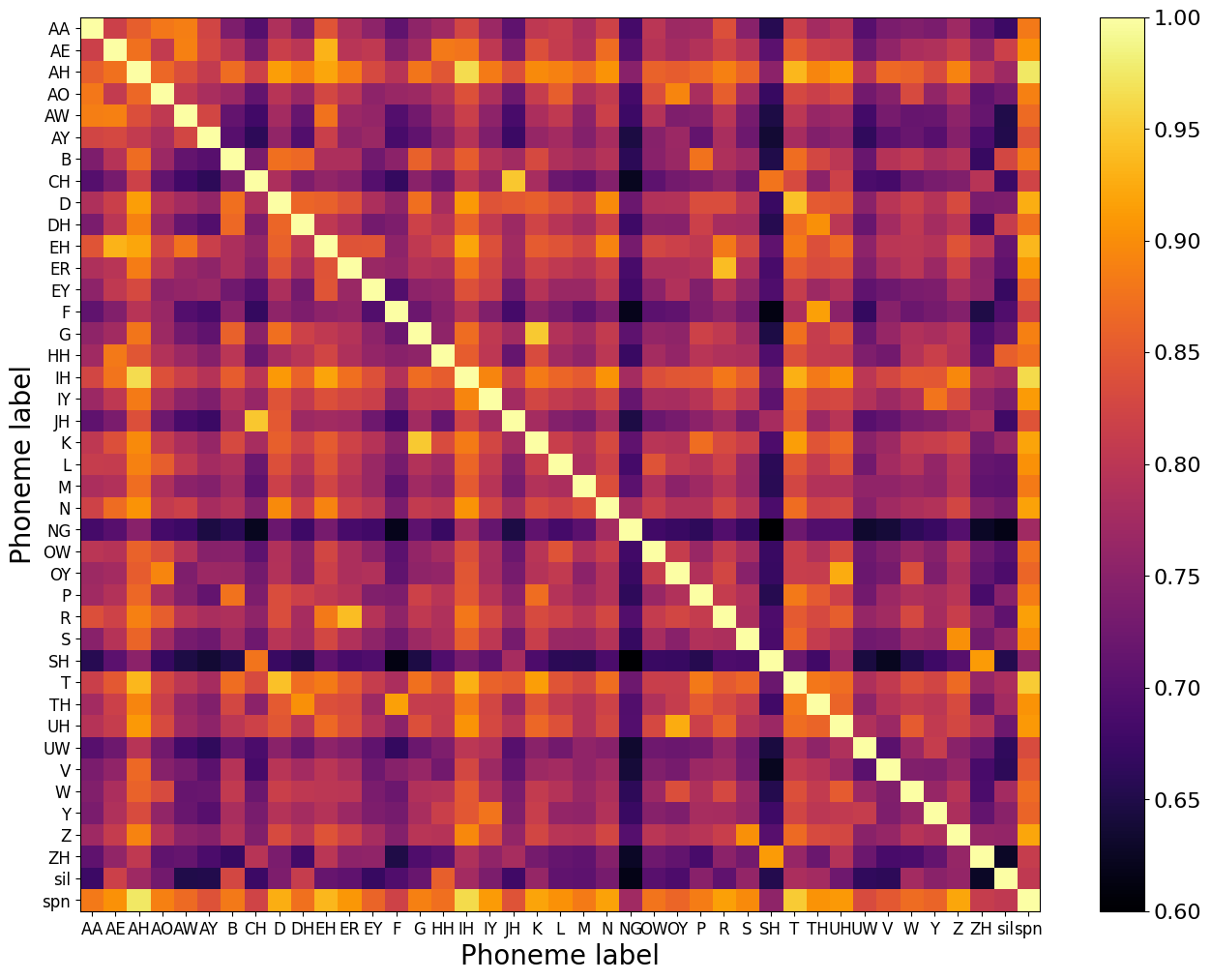}
        \caption{}
    \end{subfigure}
    \vspace{-5pt}
    \caption{(a) Average of the attention weights for each phoneme input (b) Cosine similarity between the averaged attention weights}
    \label{fig:visual}
    \vspace{-18pt}
\end{figure*}

\vspace{-12pt}
\subsection{Content embedding extraction}
\vspace{-5pt}
To change a voice's style while preserving its original content and extract speaking styles from a target speaker related to the content at each frame, we need to extract content embeddings that do not contain other style-related information than content-related.
For that, we obtain an SSL feature using a pre-trained HuBERT \cite{hubert} model\footnote{https://github.com/facebookresearch/textlesslib} and discretize the vector with the learned centroids.
The idea of utilizing a discretized SSL feature for voice conversion has been studied in previous works \cite{vqmivc, softvc} and it has been shown that it is possible to obtain content-only information from the vector quantization without the need for aligned text transcriptions.
The discretized HuBERT features are passed into a content encoder module constructed with a Transformer\cite{transformer} network to yield content embeddings.
The same content encoder is shared to extract content information from the source speech and target speech.

\vspace{-12pt}
\subsection{Style modeling with transposed dual attention}
\vspace{-6pt}
A drawback of the kNN-VC approach is that all the style-related information is scattered in the whole target set, rather than being collected in a concise, fixed-size array.
To obtain a compact style representation and extract the best style factor depending on the source speech's content, we introduce a new attention mechanism named ``transposed dual attention''.

First, a style encoder takes an encoded mel-spectrogram and the content embeddings from the target utterances to generate a sequence of embeddings.
Then, a multi-head attention (MHA) layer \cite{transformer} is applied to the content embeddings and style encoder outputs, where randomly-initialized learnable embeddings (query set) are given as queries, content embeddings are given as keys, and the style encoder outputs are given as values.
The attention layer outputs are basically the weighted sum of the style encoder outputs where attention weights are determined by the similarity between the content embedding and the query set entry.
As the size of the query set is pre-determined, the output of the attention layer, or stylebook, is also fixed in terms of the number of vectors.
It suppresses the increase of the computation cost even when the target speech set becomes large, as discussed in \cite{perceiver}.

Once the stylebook is obtained, another MHA layer is utilized to find the speaking style of the target speaker on specific content.
More specifically, the content embeddings extracted from the source speech are now given as queries and the same query set used for the stylebook generation is given as keys (i.e., the role of the query and the key is transposed).
As values, the stylebook instead of the style encoder output is given.
By doing so, we can generalize the search for a similar style from the target speech set in a non-linear and efficient way.

In order to see if the stylebook is modeling content-dependent style information, the attention weights between the source content embeddings and the learned query set, averaged for each phoneme label of the source speech are illustrated in Fig. \ref{fig:visual}(a).
To obtain the attention map, phoneme labels for each frame were determined based on the label created with Montreal-Forced-Aligner (MFA) \cite{mfa} on the LibriTTS \cite{libritts} test-clean dataset\footnote{https://github.com/kan-bayashi/LibriTTSLabel}.
From the attention map, we can see that different stylebook entries are highlighted for different input phonemes, except for a few entries that are globally used by all input.
If we measure the cosine similarity between the averaged attention weights of different phoneme labels (Fig. \ref{fig:visual}(b)), we can see that phonetically similar consonants (e.g., CH and JH, G and K) have a high similarity, which implies that the style extraction is dependent on the input's phonetic information.

Our proposed stylebook is similar to the global style tokens (GST) \cite{gst}, Attentron \cite{attentron}, and \cite{content_dependent} in that multiple style embeddings are obtained for modeling the speaking style via the attention mechanism.
However, unlike the GST which produces a global style embedding vector, the stylebook aims at obtaining content-dependent style embeddings that are necessarily different at all time frames.
Also, stylebook can greatly improve computational efficiency in extracting content-dependent information regardless of the size of target data, while previous methods \cite{knn-vc, attentron, content_dependent} suffer from increased computational time with more data.
Upon obtaining the content-dependent embeddings, our proposed model does not require any classification losses as in \cite{content_dependent}.

\vspace{-10pt}
\subsection{Speech synthesis using score-based diffusion model}
\vspace{-5pt}
We use the score-based diffusion model \cite{score_diffusion} to generate the mel-spectrogram of a converted speech.
The backbone U-Net \cite{u-net} structure is the same as in previous works\cite{diffvc, unitspeech}, but the conditioning mechanism differs in that different style embeddings are used at each frame instead of the same speaker embedding.

\begin{table*}[t]
    \caption{Objective performance of the baseline and the proposed models}
    \vspace{-8pt}
    \centering
    \begin{tabularx}{\linewidth}{c *{10}{Y}}
        \toprule
        \textbf{Metric} & \multicolumn{3}{c}{\textbf{NISQA ($\uparrow$)}} & \multicolumn{3}{c}{\textbf{SECS ($\uparrow$)}} & \multicolumn{3}{c}{\textbf{CER [\%] ($\downarrow$)}} \\
        \midrule
        \textbf{Target length} & \textbf{10 sec} & \textbf{1 min} & \textbf{5 min} & \textbf{10 sec} & \textbf{1 min} & \textbf{5 min} & \textbf{10 sec} & \textbf{1 min} & \textbf{5 min} \\
        \midrule
        \textbf{YourTTS} & 3.192 & 3.194 & 3.185 & 0.453 & 0.463 & 0.459 & 6.078 & 5.921 & 6.106 \\
        \textbf{FreeVC} & 3.574 & 3.572 & 3.563 & 0.334 & 0.347 & 0.347 & 1.895 & 1.766 & 1.831 \\
        \textbf{Diff-VC} & 3.357 & 3.343 & 3.342 & 0.414 & 0.426 & 0.426 & 5.148 & 5.372 & 5.428 \\
        \midrule
        \textbf{Proposed} (HuBERT-100) & 3.522 & 3.573 & 3.585 & 0.552 & 0.552 & 0.548 & 12.857 & 8.199 & 9.350 \\ 
        \textbf{Proposed} (HuBERT-200) & 3.681 & 3.734 & 3.744 & 0.533 & 0.538 & 0.533 & 10.900 & 8.158 & 7.882 \\ 
        \midrule
        \textbf{kNN-VC} & 3.391 & 3.564 & 3.649 & 0.612 & 0.664 & 0.674 & 8.856 & 2.511 & 1.747 \\
        \bottomrule
    \end{tabularx}
    \vspace{-15pt}
    \label{tab:experiment_result}
\end{table*}

During the training stage, the same speech given as a source is also given as a target and the model is trained to faithfully reconstruct the input speech within the autoencoder framework.
Training loss is a linear combination of a diffusion model loss and a content encoder loss,
\begin{equation}
    \mathcal{L} = \mathcal{L}_{\rm{Diff}} + \mathcal{L}_{\rm{Enc}},
\end{equation}
following the approach of \cite{unitspeech}.
In order to improve the converted speech's quality, classifier-free guidance \cite{guidance} is also introduced where the style embeddings extracted from the stylebook are randomly replaced with a trainable unconditional style embedding with the probability of $p=0.1$.

\vspace{-12pt}
\section{EXPERIMENTS}
\label{sec:experiments}
\vspace{-8pt}
\subsection{Implementation details}
\vspace{-5pt}
The structure of the content encoder is the same as the one in \cite{diffvc} and \cite{unitspeech}.
The mel encoder is a 3-layer multi-layer perceptron (MLP) and the style encoder is a 3-layer convolutional neural network (CNN) with a kernel size of 3, both with 256 hidden channels.
The query set for the stylebook includes 128 vectors each with 256 dimensions, and the dimension of the generated stylebook is reduced to 64 after a MHA layer with 2 heads and 256 embedding dimensions.
U-Net decoder structure is also the same as \cite{diffvc} and \cite{unitspeech}, where the internal base dimension is set to 128.

For the neural vocoder, we used a pre-trained HiFi-GAN \cite{hifigan} model trained on LibriTTS dataset, operating at 16 kHz sampling rate\footnote{https://huggingface.co/speechbrain/tts-hifigan-libritts-16kHz}.
As the pre-trained model takes the mel-spectrogram sampled per 256 samples while the HuBERT feature is generated per 320 samples, HuBERT features have been up-sampled with the nearest-neighbor method by the factor of 1.25.
All the mel-spectrogram in the model was generated following the pre-trained vocoder's configuration.

When obtaining the speech from the diffusion decoder, classifier-free guidance \cite{guidance} is used in the same manner as \cite{unitspeech}.
The gradient scales for the content embedding and the style embedding are set to 1.0 and 0.5, respectively. 
The number of time steps for the sampling, $N$, is set to 30.

\vspace{-15pt}
\subsection{Training scheme}
\vspace{-8pt}
For the training, two subsets from the LibriTTS dataset (train-clean-100 and train-clean-360) were used, which include 245 hours of speech from 1,151 speakers in total.
The audio files were resampled from the original 24 kHz to 16 kHz and uniformly sliced into 5-second-long segments for constructing a training batch.
Adam optimizer \cite{adam} with an initial learning rate of $10^{-4}$ was used for the parameter updates.

\vspace{-15pt}
\subsection{Evaluation methods}
\vspace{-8pt}
Three objective metrics -- NISQA \cite{nisqa}, speaker embedding cosine similarity (SECS), and character error rate (CER) -- were used to measure the performance of the models.
SECS was measured on the speaker embedding obtained from a pre-trained ECAPA-TDNN \cite{ecapa} model\footnote{https://huggingface.co/speechbrain/spkrec-ecapa-voxceleb}.
CER was measured between the source and the converted speech's text, both estimated from a pre-trained Wav2vec 2.0 \cite{wav2vec2} model\footnote{https://huggingface.co/docs/transformers/model\_doc/wav2vec2}.

For the evaluation, 28 speakers from the LibriTTS test-clean dataset have been selected and 10 pairs of source and target speeches for each speaker pair have been collected, resulting in 7,560 pairs in total.
In order to verify the performance depending on the length of the target speech, various target lengths (10 seconds, 1 minute, and 5 minutes) have been tested.
As baseline models, YourTTS \cite{yourtts}, FreeVC \cite{freevc}, Diff-VC \cite{diffvc}, and kNN-VC \cite{knn-vc} were selected and their official implementations were used to generate the samples.

\begin{table}[t]
    \caption{Memory usage for storing a target speaker's style}
    \vspace{-8pt}
    \centering
    \begin{tabularx}{\linewidth}{c *{4}{Y}}
        \toprule
        \textbf{Metric} & \multicolumn{3}{c}{\textbf{Memory usage for a target speaker ($\downarrow$)}} \\
        \midrule
        \textbf{Target length} & \textbf{10 sec} & \textbf{1 min} & \textbf{5 min} \\
        \midrule
        \textbf{YourTTS} & 2 KiB & 2 KiB & 2 KiB \\
        \textbf{FreeVC} & 1 KiB & 1 KiB & 1 KiB \\
        \textbf{Diff-VC} & 1.5 KiB & 1.5 KiB & 1.5 KiB \\
        \midrule
        \textbf{Proposed} & 32 KiB & 32 KiB & 32 KiB \\
        \midrule
        \textbf{kNN-VC} & 2,000 KiB & 12,000 KiB & 60,000 KiB \\
        \bottomrule
    \end{tabularx}
    \vspace{-18pt}
    \label{tab:memory}
\end{table}

\vspace{-15pt}
\subsection{Experimental results}
\vspace{-8pt}
Table \ref{tab:experiment_result} summarizes the objective performance of the baseline models and the proposed models with different numbers of HuBERT feature clusters (100 and 200).
In terms of speech quality reflected in the NISQA scores, the proposed model is comparable to other baseline models.
Notably, our proposed model shows a better SECS score than baseline models except for kNN-VC.
However, CER is higher than other baseline models, especially when the length of the target speech is short.
The lower CER can be due to the loss of content information from the discretization of the HuBERT feature.

Table \ref{tab:memory} shows the memory size needed to store a target speaker's speaking style, such as speaker embedding, stylebook, or WavLM features.
Baseline models using a single speaker embedding \cite{yourtts, freevc, diffvc} require a small memory size for the speaker enrollment but result in a low speaker similarity.
As can be seen from Table \ref{tab:experiment_result} and Table \ref{tab:memory}, kNN-VC performs best in SECS and CER, but only at the expense of a high memory footprint to store the target speaker's speaking style (i.e., the entire WavLM features from the target speech), which grows linearly to the target speech's length.
On the contrary, our proposed model requires a significantly smaller memory footprint than kNN-VC and it does not change depending on the target speech's length.

\vspace{-16pt}
\section{CONCLUSION}
\label{sec:conclusion}
\vspace{-10pt}
In this work, we proposed a new attention mechanism called transposed dual attention for modeling a content-dependent speaking style with a fixed-size array in an unsupervised manner.
Combining a self-supervised learning model and a stylebook with a diffusion-based model, we showed that converted speech obtained with our proposed model is closer to the target speaker than baseline models in terms of speaker similarity while providing similar speech quality.
Finally, the proposed model operates within a reasonable footprint that does not increase with target speech length.
We will further investigate this stylebook, content-dependent self-supervised style representation, in other applications that require speaking style modeling.



\vfill\pagebreak
\bibliographystyle{IEEEbib}
\bibliography{refs}

\end{document}